\documentclass{egpubl}
\usepackage{eurova2025}

\WsPaper           

\usepackage[T1]{fontenc}
\usepackage{dfadobe}  

\usepackage{cite}  
\BibtexOrBiblatex
\electronicVersion
\PrintedOrElectronic
\ifpdf \usepackage[pdftex]{graphicx} \pdfcompresslevel=9
\else \usepackage[dvips]{graphicx} \fi

\usepackage{egweblnk} 

\usepackage{wrapfig}
\usepackage{hyperref}

\usepackage{xspace}
\newcommand{\myProcessModel}[0]{{\changed{HDMI Canvas}\xspace}}

\newcommand{\changed}[1]{\textcolor{black}{#1}}

\usepackage{enumitem}

\title[\myProcessModel{}]%
      {
      \changed{The Human-Data-Model Interaction Canvas for Visual Analytics} 
      }

\author[Jürgen Bernard]
{\parbox{\textwidth}{\centering Jürgen Bernard$^{1,2}$\orcid{0000-0001-8741-9709} 
        }
        \\
{\parbox{\textwidth}{\centering $^1$University of Zurich, Switzerland; $^2$Digital Society Initiative, Zürich, Switzerland \\
       }
}
}

\begin{document}

\teaser{
 \vspace{-8mm}
 \includegraphics[width=0.9\linewidth]{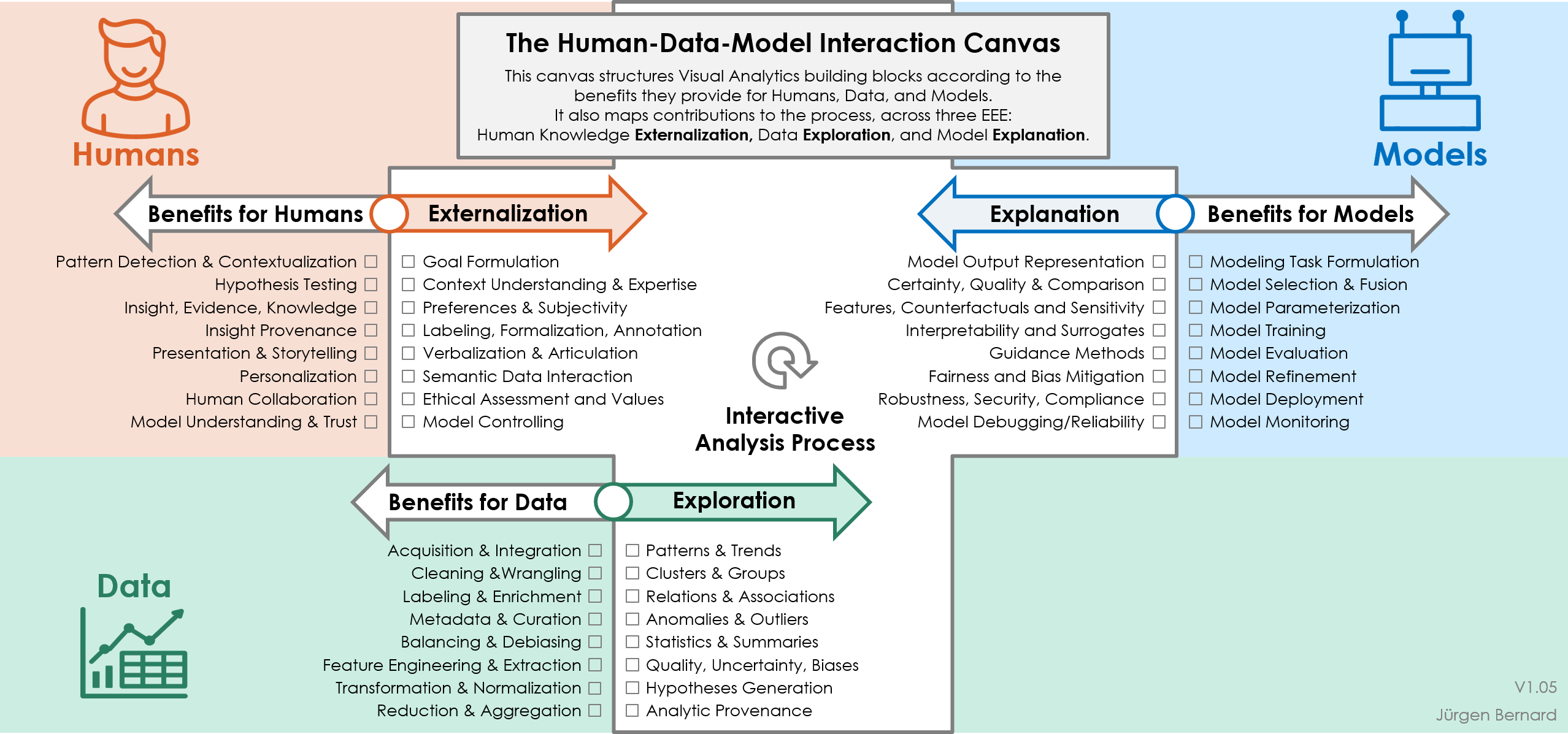}
 \centering
  \caption{\changed{The \myProcessModel{} defines six lists of key VA building blocks that describe how humans, data, and models contribute to the VA process and benefit from the process. The canvas supports the characterization of existing approaches and the design of novel VA solutions}.}
\label{fig:teaser}
}

\maketitle
\begin{abstract}
   Visual Analytics (VA) integrates humans, \changed{data, and models} as key 
   actors in insight generation and data-driven decision-making.
   This position paper values and reflects on \changed{16} VA process models and frameworks and makes \changed{nine} high-level observations that motivate a fresh perspective on VA.
   The contribution is the \myProcessModel{}, a perspective to VA that complements the strengths of existing VA process models and frameworks.
   It systematically characterizes diverse roles of humans, \changed{data, and models,} and how these \changed{actors} benefit from and contribute to VA processes.
   The descriptive power of the \myProcessModel{} eases the differentiation between a series of VA building blocks, rather than describing general VA principles only.
   The canvas includes modern human-centered methodologies, including human knowledge externalization and forms of feedback loops, while interpretable and explainable AI highlight model contributions beyond their conventional outputs.
   The \myProcessModel{} has generative power, guiding the design of new VA processes 
   \changed{and is optimized for external stakeholders, improving VA outreach, interdisciplinary collaboration, and user-centered design.}
   The utility of the \myProcessModel{} is demonstrated through two preliminary case studies.
   


\printccsdesc   
\end{abstract}  
\section{Introduction}
\label{sec:introduction}

The evolution of Visual Analytics (VA) has been an exciting journey.
Since its formal characterization as a successor of visual data mining about 20 years ago, VA has seen constant change, shaped by a series of external developments and trends like \changed{advanced statistics}, big data, data science, \changed{(interactive)} machine learning (ML), deep learning, large language models, and AI in general.
Today, the three \changed{actors} \textit{Humans}, \changed{\textit{Data}, and} \textit{Models} (including statistics, analysis algorithms, and ML) are more prominently discussed in society than ever before, signaling a prosperous era for VA research and practice, leading to enhanced data-, visualization-, interaction-, and AI-literacy.
\changed{In the era of the digital transformation and data-driven decision-making, VA provides practical solutions to pressing challenge areas and emerging trends, including data exploration,
scalability, complexity handling,
multi-criteria decision-making,
explainability, interpretability, transparency,
bridging semantic gaps,
domain-specific adaptability,
cognitive amplification}, and not least: human-AI collaboration.

Questions \changed{that inspired} this position statement are: How well does our community sell its unique assets?
Does the world know?
A latent concern is that VA undersells its immense benefits to the outside world.
Or, more constructively: How can we enhance the visibility and adoption of VA assets?
Which VA principles need to be included in a one-minute pitch to a real-world stakeholder? 
To what extent do VA research agendas, process models, and theoretical frameworks convey these key principles to society \changed{at first sight}? 
Reflections within the VA community and experiences from applied, collaborative work have surfaced \changed{nine} key observations:

\textbf{Knowledge Generation beats Knowledge Externalization}: 
Traditional process models in data mining, KDD, ML, and VA prioritize knowledge generation, supporting the flow from data to insights~\cite{fayyad1996data,crispdm1999,wirth2000crisp,pirolli2005sensemaking,cook2005illuminating,Keim2008,sachaKGM2014,endertRTWNBR17}. 
However, the human knowledge externalization process \changed{(from tacit~\cite{1995_Noanaka_Takeuchi} into externalized knowledge through interactive interfaces)
seems to be much less reflected.} 
Deeply rooted in social sciences~\cite{militello1998applied,koh2015tools,klein1989critical,chipman2000introduction}, \changed{knowledge externalization was proposed for VA} by Wang et al. and Federico et al.~\cite{wangJDLRC09,federico2017,ceneda2016characterizing}, leveraging van Wijk's influential value-of-visualization model.
\changed{A fresh perspective may balance knowledge generation and externalization}.

\textbf{Feedback Loop++}: The feedback loop~\cite{Keim2008,sachaKGM2014,sachaSZLWNK16,sachaSZLWNK16} is one of VA's assets, beyond less human-centered \changed{Knowledge Discovery in Databases (KDD)}~\cite{fayyad1996data} or \changed{Cross-Industry Standard Process for Data Mining (CRISP-DM)}~\cite{crispdm1999,wirth2000crisp} processes.
Initially focused on model selection, parameter steering, and quality assessment, human feedback now encompasses broader elements~\cite{endertHRNFA14,amershi2014power,holterE24}.
With the ''human is the loop'' scope~\cite{endertHRNFA14}, humans can influence similarities, dimension weights, item weights and relevance, group counts and contents, centroid landmarks, and labels.
Additionally, distinctions exist between core knowledge and alternatives like domain expertise, contextual understanding, preferences, subjectivity, emotions, ethical assessment, and values~\cite{amershi2014power, holterE24}. 
\changed{Modern perspectives also ask about \textit{how} feedback is elicited and formalized}, including structured feedback~\cite{wscg2014,bernardcgf2018}, intervention~\cite{fails2003interactive,amershi2014power,endertHRNFA14}, semantic interaction~\cite{endertFN12,endertFN12Text}, human-data interaction~\cite{mortierHHMC14}, and interactive adaptation~\cite{ramos2020interactive,wondimu2022}.

\textbf{Role of the Human}: Many process models are highly abstract, or humans are only indirectly represented through visualization or knowledge components~\cite{fayyad1996data,crispdm1999,wirth2000crisp,pirolli2005sensemaking,Keim2008}, obscuring the full spectrum of human involvement.
A fresh, human-centered perspective may explicitly distinguish how \changed{humans \textit{benefit} from VA processes~\cite{riedl2019,shneiderman21,shneiderman22,russell2022} and how they can \textit{contribute}} actively~\cite{ramos2020interactive,wondimu2022,amershi2014power}.
This approach can prioritize needs, expertise, and interpretability, with feedback mechanisms ensuring that processes remain effective, ethical, and aligned with real-world needs~\cite{shneiderman21}.

\textbf{Role of the AI}: AI models have evolved from a passive tool that require manual selection, parameterization, and human oversight~\cite{fayyad1996data,crispdm1999,wirth2000crisp,pirolli2005sensemaking,Keim2008,sachaKGM2014,sachaSZLWNK16,sachaSZLPWNK17,sachaKKC19}, to active (mixed-initiative) agents that more autonomously analyze data, learn, adapt, and collaborate with humans in decision-making~\cite{allen1999mixed,horvitz99,amershi2014power,riedl2019,cag2021sperrle,shneiderman21,shneiderman22,russell2022}. 
\changed{Treating models as actors}, including their needs for improvement, but also their explainability, interpretability, and reliability~\cite{spinnerSSE20,hohman21,chatzimparmpasM20,rosaBBASCBGA23} \changed{would align VA with human-model collaboration perspectives~\cite{rezwanaM23,fragiadakis2024,holterE24}.}

\textbf{Role of the Data}: 
\changed{Data was long seen as a passive repository of gathered information that was processed, mined, and exploited} for decision-making~\cite{pirolli2005sensemaking,kitchin2014data}.
However, recent frameworks reconceptualize data as socially constructed artifacts embedded with human values, biases, and context~\cite{d2023data, lupi2017data, lin2022data}. 
Rather than merely processed algorithmically with AutoML~\cite{rogersABCFKKKTW24}, data is increasingly understood to be engaged with both analytically and intuitively~\cite{mortierHHMC14}, inspiring exploration, insight formulation, developing ''data hunches''~\cite{lin2022data}, and ''spending time with data''~\cite{lupi2017data}. 
\changed{This VA perspective aligns with \textit{data feminism}, exposing power imbalances in data~\cite{d2023data}; \textit{data humanism}, emphasizing emotional dimensions of data~\cite{lupi2017data}; and \textit{personalized analytics}, prioritizing individual meaning-making over standardized metrics~\cite{mortierHHMC14}}. 

\textbf{Goal(s) of a VA Process}:
A prominent perspective of VA is as a flow from data to knowledge, using models, visualizations, and interactions as tools~\cite{fayyad1996data,crispdm1999,card1999,pirolli2005sensemaking,cook2005illuminating,Keim2008,sachaKGM2014}.
However, does this always match real-world scenarios~\cite{oppermannM20}?
What if, in some cases, data and knowledge are provided, so the goal is to benefit model enhancement~\cite{sachaKKC19}? 
Alternatively, knowledge and models may exist, yet data is lacking ''AI-readiness''.
\changed{Like VIAL for interactive data labeling~\cite{euroVA2017Labeling,bernardZSA18}, VA may describe three simultaneous goals: benefiting human knowledge, model improvement, and data enrichment.}

\textbf{Three Types of Processes}:
In VA, three process models dominate. 
i) \textit{Data transformation processes}, rooted in data mining and ML~\cite{fayyad1996data,crispdm1999,wirth2000crisp,card1999,Keim2008,sachaSZLWNK16,sachaSZLPWNK17,hohman21}, prepare data and ease pattern detection, executed mainly by developers. 
ii) \textit{Knowledge generation processes}, emerging from information seeking and sense-making~\cite{marchionini1995information,shneiderman96,pirolli2005sensemaking,card1999,kehrerLMDSH08,Keim2008,sachaKGM2014}, guide users in decision-making. 
iii) \textit{Design processes}, mainly conducted by visualization designers~\cite{munzner09,sedlmairMM12,oppermannM20,meyerD20}, although VA-specific support remains scarce.
\changed{A fresh perspective may present VA to the outside world in a unified framework, leveraging principles from data transformation, knowledge generation, and design.}

\textbf{Descriptive Power}: 
\changed{Conceptual structures} with descriptive power help to distinguish approaches, helping people to think--within the community and ideally beyond.
Many VA process models outline essential characteristics and fundamental components \changed{that all VA approaches share}~\cite{fayyad1996data,crispdm1999,card1999,pirolli2005sensemaking,Keim2008,meyerD20,sachaKGM2014}, much like a software interface or abstract base class.
However, they do not necessarily help to differentiate between approaches \changed{and their VA building blocks}. 
\changed{Enhanced perspectives may incorporate more detailed structure about human, data, and model factors to ease the characterization of applications}.

\textbf{Generative Power}: 
\changed{Designing VA systems is even more challenging than traditional visualization systems, where methodological design support is more prevalent~\cite{munzner09,sedlmairMM12}.}
\changed{The generative power of traditional VA process models sometimes falls short in characterizing problems, identifying essential building blocks, and enabling the generation of novel, non-trivial VA solutions within a structured design space.}
\changed{A detailed VA design space could address these gaps and enhance interdisciplinary collaboration with external stakeholders in participatory and user-centered design.}

\vspace{3mm}
The contributions of this position paper are as follows.
\begin{itemize}
    \item It summarizes and values the strengths of \changed{16} VA process models, conceptual frameworks, and design spaces, while identifying potential extensions, provided in the Supplemental Materials. Used to substantiate the \changed{nine} observations presented in Section~\ref{sec:introduction}.
    \item \changed{It introduces a perspective on \textit{Humans}, \textit{Data}, and \textit{Models}, as \textit{actors} (as used in socio-technical contexts), to highlight their participation and interaction. The perspective analyzes how actors \textit{benefit from} and \textit{contribute to} VA processes, using the metaphor of three ingoing and three outgoing arrows, to show interaction directions. The three types of contributions to the VA process form the \textit{EEE Framework}: \textit{Externalization} of human knowledge, \textit{Exploration} of data, and \textit{Explanation} of models (Section~\ref{sec:contributeBenefit})}.
    \item \changed{It presents the \myProcessModel{} (Figure~\ref{fig:teaser}), an assembly of humans, data, and models, with $3 \times 2$ directed arrows representing \changed{actor} contributions and \changed{actor} benefits.
    For each of the six arrows, the \myProcessModel{} defines key VA building blocks, enabling both the characterization of existing approaches and the design of novel VA approaches. 
    The \myProcessModel{} helps to break down the complexity of an analysis problem and to reveal human-data-model collaboration patterns, using a language suitable for external stakeholders (Section~\ref{sec:canvas}).
    \item \changed{First preliminary} case studies demonstrate the practical application of the \myProcessModel{} in concrete VA systems} (Section~\ref{sec:cases}).
    \item \changed{An online \href{https://forms.gle/ysEj1PrELrra7qM37}{feedback mechanism} before, during, and after the EuroVA workshop, to collect detailed feedback, validate the canvas structure, and ensure its adoption in the VA community}.
\end{itemize}

\section{Related \changed{Process Models, Frameworks, and Design Spaces}}
In addition to the \changed{nine} high-level observations presented in the introduction that motivate the \myProcessModel{}, supplemental material evaluates the strengths and remaining opportunities of \changed{16} VA process models, conceptual frameworks, \changed{and design spaces}--for reflection and for forming the fresh perspective.



\section{\changed{Roles of Humans, Data, and Models in VA Processes}}
\label{sec:contributeBenefit}

\changed{
VA processes center around three key \changed{actors}: Humans, Data, and Models.
This section discusses how these \changed{actors} \textit{benefit} from VA (Section~\ref{sec:canvas:benefits}) and how they \textit{contribute} back to the VA process (Section~\ref{sec:canvas:contributions}).
The notion of benefits and contributions relates to traditional goal and task-based framings, where the VA process is tasked to benefit the actors, and actors are tasked to contribute to the VA process.
Beyond tasks, benefits for and contributions by actors emphasize the bilateral nature of sending and receiving, as fundamental forms of communication. 
A main contribution is the "EEE" Framework, combining the \textbf{Externalization} of human knowledge, the \textbf{Exploration} of Data, and the \textbf{Explanation} of algorithmic models.
This perspective supports the study of human-data-model collaboration, laying the ground for the \myProcessModel{} in Section~\ref{sec:canvas}.
}

\subsection{How Actors Benefit from Visual Analytics Processes}
\label{sec:canvas:benefits}
\setlength{\columnsep}{8pt}%
\begin{wrapfigure}[7]{r}{0.37\linewidth}  
    \centering
    \vspace{-10pt}
    \includegraphics[width=\linewidth]{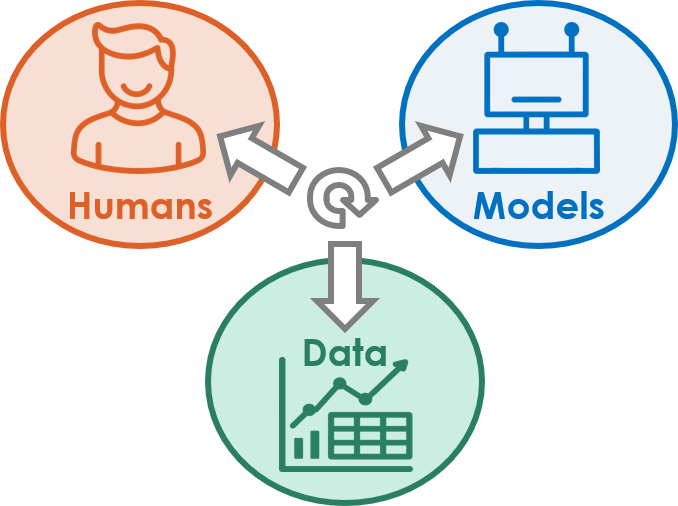}
\end{wrapfigure}
The benefit describes the ''centeredness'' of VA processes for \changed{actors}, i.e., what they take away.
\changed{In human-centered VA, the output is knowledge, as a human benefit.
Data-centric approaches enhance the data itself, while in ML, models benefit from VA processes}. 

\textbf{Human-centered} 
VA processes enhance data-driven decision-making by empowering users to explore complex data, derive meaningful insights, and make informed choices.
Aligned with human-centered AI, and in contrast to fully automated approaches, human-centered VA prioritizes user involvement, interpretability, and adaptability. 
This ensures that data analysis remains intuitive, transparent, and actionable.
Key benefits of this approach include:

\begin{itemize}
    \item Pattern detection and data structuring
    \item Pattern contextualization of observations and findings
    \item Confirmatory data analysis to accept or reject hypotheses
    \item Insight, evidence, and knowledge generation
    \item \changed{Presentation, reporting, and storytelling}
    \item Personalization of the analysis and insight provenance
    \item Collaboration and shared exploration among users
    \item \changed{Model understanding and trust}
\end{itemize}

\textbf{Data-centered} VA processes help to transform raw data into representations that are usable by \changed{algorithmic models} and useful for downstream analyses. 
Data can capture user interactions, feedback, and contextual nuances, enabling both humans and models to contribute to data improvement.
Key benefits for data include:
\begin{itemize}
    \item Data acquisition, integration, and connectivity
    \item Data cleaning and wrangling
    \item Data labeling, augmentation, and contextual enrichment
    \item \changed{Data curation and metadata}
    \item \changed{Data balancing and debasing}
    \item \changed{Data feature extraction and engineering}
    \item Data transformations and normalizations
    \item \changed{Data reduction, and aggregation}
\end{itemize}

\textbf{Model-centered} VA aligns with interactive ML: Based on a task formulation and ''AI-ready'' data, model-centered processes focus on parameterizing, building, refining, evaluating, deploying, and monitoring models. 
Notable is the need for humans in essential process steps, e.g., for data labeling in supervised ML.
Models benefit from human oversight, feedback, and contextual understanding, which help improve their accuracy, relevance, and ethical alignment.
Benefits for models through VA can be described as:
\begin{itemize}
    \item \changed{Modeling task formulation}
    \item Model selection, model fusion, and ensembles
    \item Model parameterization
    \item \changed{Model building and training}
    \item \changed{Model refinement}
    \item Model evaluation and quality assessment
    \item Model deployment, monitoring, and performance tracking
\end{itemize}

\subsection{How Entities Contribute to VA -- The EEE Framework}
\label{sec:canvas:contributions}
\setlength{\columnsep}{8pt}%
\begin{wrapfigure}[8]{r}{0.365\linewidth}  
    \centering
    \vspace{-10pt}
    \includegraphics[width=\linewidth]{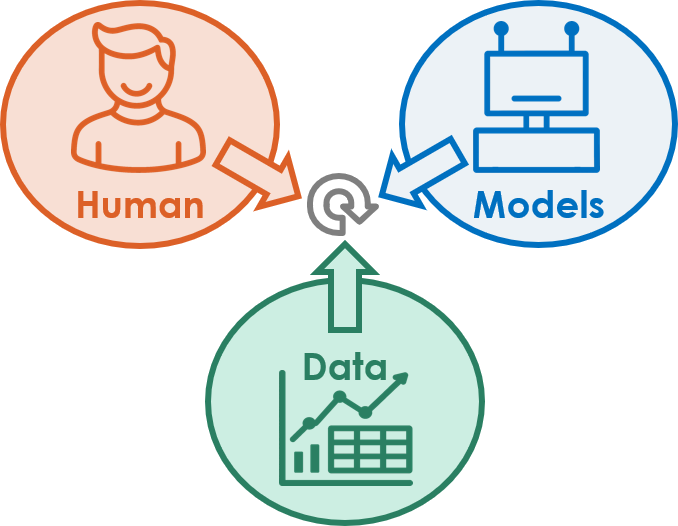}
\end{wrapfigure}
VA processes can be structured by the contributions of the three key \changed{actors}. 
Each \changed{actor} brings unique strengths, enabling robust control, communication, and mutual guidance, forming the ''\textbf{EEE Framework}'': \changed{Human Knowledge and Preference \textbf{Externalization}, Data \textbf{Exploration}, and Model \textbf{Explanation}}.

\textbf{(Human Knowledge and Preference) Externalization} transforms internal, tacit expertise into explicit, communicable knowledge or preferences. 
Capturing expert knowledge, expertise, feedback, and preferences tailors data analysis for strategic and/or personalized decision-making. 
Well-established in cognitive science, integrating externalization and interactive elicitation practices in VA can strengthen methodological support for data analytics:
\begin{itemize}
    \item \changed{Goal formulation}
    \item Knowledge and contextual understanding
    \item Domain expertise and experiential knowledge 
    \item Preferences, subjectivity, and emotions
    \item Interactive data labeling, formalizing knowledge and preferences
    \item Annotation support and data hunches to capture tacit insights
    \item Verbalization and NLP interfaces enable user articulation
    \item Semantic interaction interfaces for direct human-data interaction 
    \item \changed{Visual query systems and advanced facets for search and filtering}
    \item Ethical assessment, creativity, and human values
    \item User control for modeling, parameterization, and validation
    \item Human collaboration environments for shared decision-making
    
\end{itemize}

\textbf{(Data) Exploration} 
is fundamental for actors in data analysis for knowledge generation, data-driven decision-making, and informed model-building. 
VA integrates human and model strengths to facilitate data discovery, navigating high-dimensional spaces using sophisticated VA methods to uncover latent structures and patterns. 
Key contributions of VA-driven data exploration include:

\begin{itemize}
    \item Representation of patterns and trends in high-dimensional data
    \item Detection of clusters and groups of data items
    \item Relation and association discovery between attributes
    \item Detection of anomalies and outliers
    \item Statistics, summaries, and structuring 
    \item Detection of quality issues, uncertainties, and biases
    \item Hypothesis generation based on explored insights
    \item Insight-based modeling and analytic provenance
\end{itemize}

\textbf{(Model) Explanation} goes beyond computational model outputs. 
Its contributions involve activities designed to understand the reasoning behind model decisions, i.e., not only receiving what output has been generated by a model, but also why and how.
It enhances interpretability through visualization and explainable AI methods, to foster a trustworthy, reliable, and safe AI experience.
Main contributions of models and model explanations include:

\begin{itemize}
    \item Model output visualization and interpretation
    \item Model certainty, quality assessment, and model comparison
    \item Guidance methods
    \item Feature importance and example-based explanations
    \item Model interpretability and surrogate models 
    \item Counterfactuals, what-if scenarios, and sensitivity 
    \item Visual, interactive, and verbal explanations 
    \item Fairness and bias mitigation methods 
    \item Robustness, security, and regulatory compliance 
    \item Improved model debugging and continuous monitoring
\end{itemize}

\section{The Interactive Visual Data Analysis (IVDA) Canvas}
\label{sec:canvas}

The \myProcessModel{} characterizes \textit{Humans}, \textit{Data}, and \textit{Models} based on the \textit{benefits} they gain from VA processes and their \textit{contributions} to the VA process.
Section~\ref{sec:canvas:requirements} discusses requirements, Section~\ref{sec:canvas:overview} presents the \myProcessModel{}, and Section~\ref{sec:canvas:usage} discusses its use. 

\subsection{Requirements}
\label{sec:canvas:requirements}

The reflection on \changed{VA observations, process models, and other conceptual structures} reveals key requirements for a fresh perspective.
\begin{itemize}
    \item Describe VA for external stakeholders with \changed{actors}, connections, and terminology that aid understanding and ease outreach. 
    \item Highlight its core \changed{actors} \textit{Humans}, \textit{Models}, and \textit{Data} and their active roles: humans as decision-makers, models as collaborative actors, and data in an activated form to underline its importance.
    \item Associate \changed{actors} with their \textit{contributions} to and \textit{benefits} from VA process to highlight interaction directions. 
    \item Extend the diverse forms of human knowledge externalization to emphasize feedback-loop beauty.
    \item Make interpretable and explainable AI methods more present as model contributions (why), beyond model output (what).
    \item Focus on identifying differences between VA approaches rather than just commonalities, to strengthen its descriptive power.
    \item Visible forms of human-model communication should characterize mixed-initiative and human-AI collaboration methods.
    \item Aid the design of new approaches through enhanced \changed{structural} guidance on VA building blocks, increasing its generative power.
    \item Visually, it should prioritize clarity, holistic and systematic thinking, and broad applicability beyond the VA community. 
\end{itemize}
To achieve this, a \textit{canvas} is used as a powerful strategic tool for structuring thoughts, analyzing situations, and supporting decision-making. 
Inspiration comes from established models like the SWOT~\cite{gurel2017swot} analysis matrix, the Business Model Canvas (BMC)~\cite{osterwalder2010business}, and the Value Proposition Canvas~\cite{osterwalder2014value}.

\subsection{Interactive Visual Data Analysis Canvas Overview} 
\label{sec:canvas:overview}

Figure~\ref{fig:teaser} introduces the \myProcessModel{}, a canvas structure that positions \textit{Humans}, \textit{Models}, and \textit{Data} based on their \textit{benefits} from VA processes (outgoing arrows) and \textit{contributions} (ingoing arrows).
It represents these \changed{actors as key elements} in the \changed{canvas}, focusing on their roles as key information hubs. 
The canvas highlights bidirectional information flow, characterizing forms of communication between \changed{actors}, fostering control, interaction, and mutual guidance.
By emphasizing dynamic interplay between \changed{actors}, it reflects continuous dialogues and learning between \changed{actors}, opening new avenues for human-\changed{data-model} collaboration.

\subsection{\myProcessModel{} Usage with External Stakeholders}
\label{sec:canvas:usage}

\changed{The \myProcessModel{} supports communication with external stakeholders by making VA building blocks and interaction flows accessible to audiences unfamiliar with VA terminology.
Many terms used in the canvas are already established in broader public and professional discourse, especially in the AI context.
This allows practitioners, data scientists, and domain experts to engage with VA concepts without prior specialized knowledge.
The structural depth of the \myProcessModel{} enables transparent descriptions of VA approaches. 
The \myProcessModel{} also serves as a generative tool for VA system design and supports user-centered development by offering a shared structure for discussing system goals and interaction-flow design.
When all $3 \times 2$ flows are relevant, the design goal is a fully integrated VA “Supertool” for human-data-model collaboration.
In summary, it provides a structured and an inclusive way to describe, analyze, and design VA systems, bridging disciplinary gaps and supporting effective collaboration with external stakeholders.
}

\section{Case Studies}
\label{sec:cases}

\begin{figure}[t]
    \vspace{-2mm}
    \includegraphics[width=\linewidth]{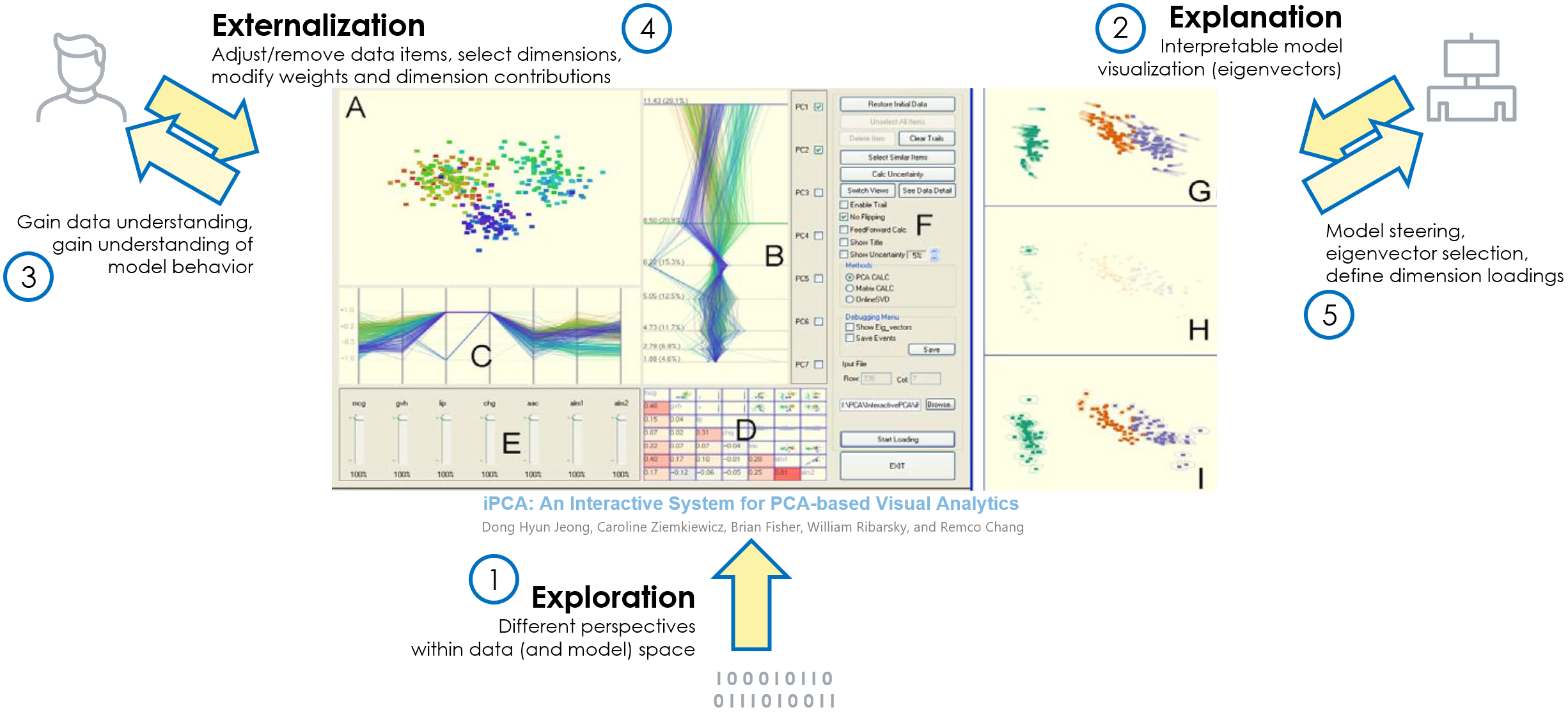}
    \caption{The iPCA~\cite{jeongZFRC09} white box approach leverages five out of six key information flows to enable \changed{performing} both exploratory data analysis and model analysis (PCA).}
    \label{fig:iPCA}
    \vspace{-6mm}
\end{figure}

\subsection{iPCA: Opening the Black Box of PCA Projections}
The iPCA~\cite{jeongZFRC09} VA approach enables experts to \changed{perform} both exploratory data analysis and model analysis (Figure~\ref{fig:iPCA}).
It demonstrates white box model-steering and interpretation for the linear principal component analysis (PCA) projection, \changed{supporting better understanding and use}.
To facilitate data exploration, iPCA visualizes \changed{PCA results, using} multiple coordinated views and a rich set of user interactions (1) (2). 
\changed{This helps experts to understand} data characteristics and relationships between the data and the calculated PCA eigenspace, leading to a better understanding of the model and its internal analytics behavior (3).
From an externalization perspective, several interaction techniques enable experts to engage with data values and model characteristics, including adjusting/removing data items, selecting dimensions, and modifying weights and dimension contributions (4).
Model-centered interactions lead to model steering through parameterization, weighting, and dimension contributions (5); direct feedback reflects changes back to the user.
This allows users to explore how and why PCA results \changed{respond} to different inputs \changed{supporting more accurate analysis and improved model transparency}. 
Overall, iPCA offers human, model, and data contributions to the VA process, with benefits for humans and models, i.e., five out of six key information flows are supported with this pioneering white box approach.

\subsection{Analyzing Correlation Patterns of Electrical Engines}

\begin{figure}[t]
    \vspace{-2mm}
    \includegraphics[width=\linewidth]{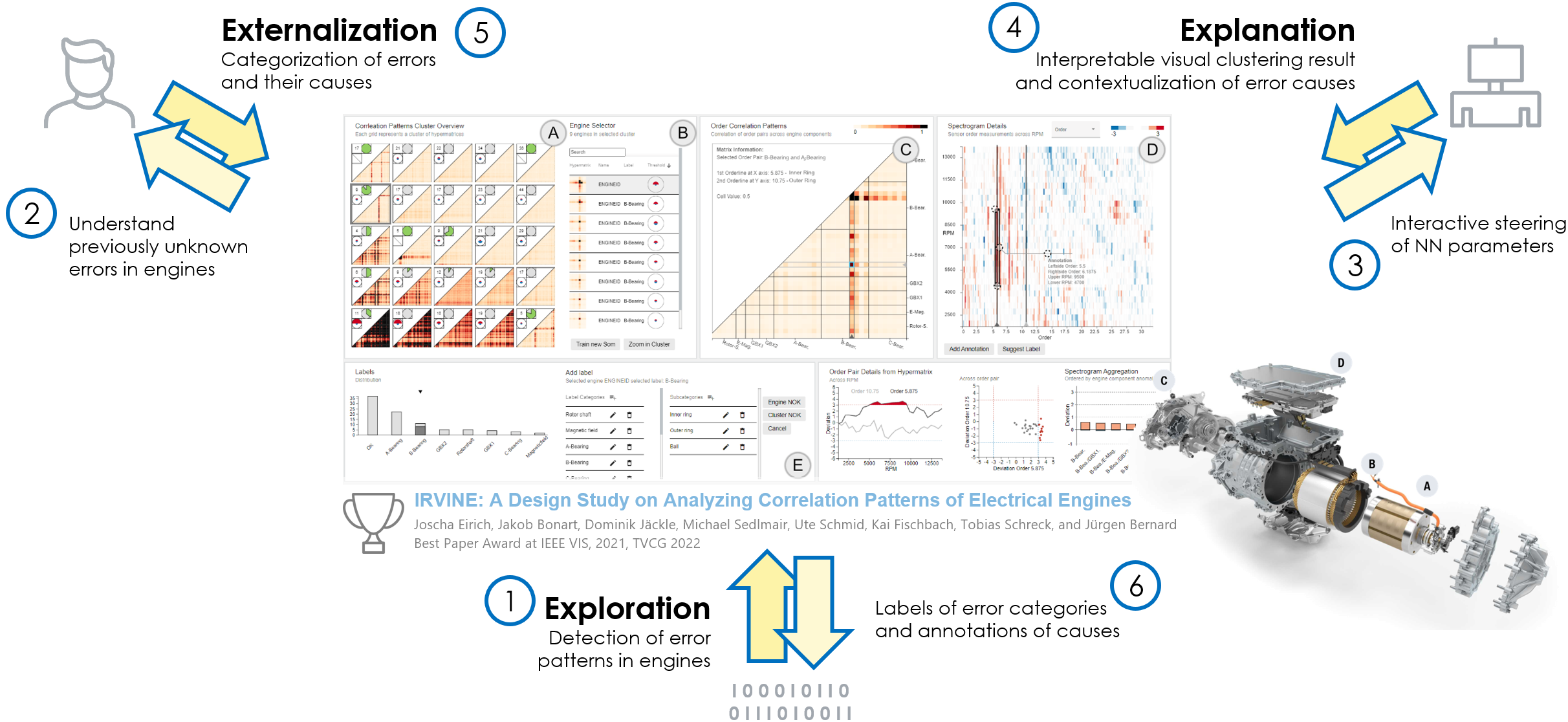}
    \caption{The IRVINE~\cite{tvcg2022eirich} VA system has ''Supertool'' complexity: to solve the analysis goal of the experts, all three \changed{actors} want to benefit from and contribute to the VA workflow (steps 1-6).} 
    \label{fig:irvine}
    \vspace{-6mm}
\end{figure}

BMW uses acoustic data analysis to improve the manufacturing of its electrical engines by detecting and understanding previously unknown errors. 
With IRVINE~\cite{tvcg2022eirich}, expert engineers apply a comprehensive data exploration (1) strategy to identify subtle error patterns (2) across engine components (Figure~\ref{fig:irvine}). 
An interactive visual clustering algorithm, a neural network-based approach, allows dynamic paramete fine-tuning (3). 
Interactive steering enhances anomaly detection and facilitates a deeper understanding of underlying issues.
The model explanation component builds trust by providing interpretable visual representations of clustering results and contextualizing potential error causes, easing insight generation (4). 
Experts engage in knowledge externalization by systematically categorizing errors and annotating potential causes (5). 
This process enriches the underlying data repository by contributing well-defined labels for error categories (6), which in turn strengthens future analysis and diagnostics. 
Overall, this integrated approach empowers BMW’s experts to transform raw acoustic data into actionable intelligence, improving engine manufacturing and maintenance strategies. 
This innovative process not only mitigates production risks but also enhances overall product quality.

\section{Discussion and Limitations}

\textbf {What is Knowledge?}: 
Unlike many VA process models and conceptual structures, the \myProcessModel{} does not treat knowledge as a main entity. 
First, because humans are an entity directly.
Second, because knowledge is not the only human value worth \changed{considering}.
Third, because defining knowledge is difficult, what information artifacts it includes, and what not.
Depending on the \changed{research context, different notions exist}, such as in strong connection to epistemology in philosophy, or evidence in empirical research.
The \myProcessModel{} uses knowledge and preferences as a tandem to highlight the need for both objective and subjective information.

\textbf{Data Actors?}:
The role of data as an independent actor in VA processes remains debated. 
At its most passive, data is merely an information state.
At its most active, it interacts with the VA system, receiving inputs and generating outputs—like a database responding to a query (''bring the algorithms to the data'').
For designers and developers, key questions arise: 
What responsibilities would data actors have in a VA process?
How should data interact with humans and models? 
What goals should it pursue? 
To function meaningfully, what must it receive from human and model actors, and what does data contribute to the overall VA objectives?
These reflections challenge the passive perception of data, opening avenues for treating it as a more dynamic \changed{actor}--be it in the design process or during data analysis--an idea for future work.

\textbf{VA Outreach}:
Part of the goal was to create a canvas that also helps to communicate essential building blocks of powerful VA systems to external stakeholders, using a form of communication that helps external people to on-board easily, without the need to internalize VA-intrinsic methodologies.
For that purpose, the \myProcessModel{} is already in use, and qualitative feedback from external stakeholders is positive, indicating its utility, and valuing VA as a central research field in the human-model-data spectrum.
However, empirical data does not yet exist to back up the utility claim made.

\section{Conclusions}

This position paper reflects on VA process models and community observations, motivating a fresh perspective on VA, useful for collaboration with external stakeholders.
Beyond the scope already nicely covered by state-of-the-art, it introduces the \myProcessModel{}, which organizes humans, data, and models as actors in a VA canvas. 
The \myProcessModel{} provides enhanced \changed{structural depth}, clarifying how each of the \changed{actors} can benefit from and contribute to VA processes more collaboratively.
The descriptive power of the \myProcessModel{} helps to distinguish existing approaches as demonstrated in two case studies, while the generative power can aid the design of novel VA approaches in collaborative and interdisciplinary settings.

\bibliographystyle{eg-alpha}  
\bibliography{main}        

\newcommand{\etalchar}[1]{$^{#1}$}
\begin{thebibliography}{\uppercase{ACKK14}}

\bibitem[ACKK14]{amershi2014power}
\textsc{Amershi S., Cakmak M., Knox W.~B., Kulesza T.}:
\newblock Power to the people: The role of humans in interactive machine learning.
\newblock \emph{AI Magazine 35}, 4 (2014), 105--120.

\bibitem[AGH99]{allen1999mixed}
\textsc{Allen J.~E., Guinn C.~I., Horvtz E.}:
\newblock Mixed-initiative interaction.
\newblock \emph{IEEE Intelligent Systems and their Applications 14}, 5 (1999), 14--23.

\bibitem[BSR{\etalchar{*}}14]{wscg2014}
\textsc{Bernard J., Sessler D., Ruppert T., Davey J., Kuijper A., Kohlhammer J.}:
\newblock User-based visual-interactive similarity definition for mixed data objects-concept and first implementation.
\newblock In \emph{Proceedings of WSCG} (2014), vol.~22, Eurographics, Vaclav Skala - Union Agency, pp.~329--338.

\bibitem[BZL{\etalchar{*}}18]{bernardcgf2018}
\textsc{Bernard J., Zeppelzauer M., Lehmann M., M\"{u}ller M., Sedlmair M.}:
\newblock Towards user-centered active learning algorithms.
\newblock \emph{Computer Graphics Forum (CGF)} (2018), 121--132.

\bibitem[BZSA17]{euroVA2017Labeling}
\textsc{Bernard J., Zeppelzauer M., Sedlmair M., Aigner W.}:
\newblock {A Unified Process for Visual-Interactive Labeling}.
\newblock In \emph{EuroVis Wkshp. Visual Analytics (EuroVA)} (2017), Eurographics.

\bibitem[BZSA18]{bernardZSA18}
\textsc{Bernard J., Zeppelzauer M., Sedlmair M., Aigner W.}:
\newblock {VIAL:} a unified process for visual interactive labeling.
\newblock \emph{Vis. Comput. 34}, 9 (2018), 1189--1207.

\bibitem[CCK{\etalchar{*}}99]{crispdm1999}
\textsc{Chapman P., Clinton J., Kerber R., Khabaza T., Reinartz T., Shearer C., Wirth R.}:
\newblock \emph{CRISP-DM 1.0: Step-by-step Data Mining Guide}.
\newblock Tech. rep., CRISP-DM Consortium, 1999.

\bibitem[CGM{\etalchar{*}}16]{ceneda2016characterizing}
\textsc{Ceneda D., Gschwandtner T., May T., Miksch S., Schulz H.-J., Streit M., Tominski C.}:
\newblock Characterizing guidance in visual analytics.
\newblock \emph{IEEE transactions on visualization and computer graphics 23}, 1 (2016), 111--120.

\bibitem[CMJ{\etalchar{*}}20]{chatzimparmpasM20}
\textsc{Chatzimparmpas A., Martins R.~M., Jusufi I., Kucher K., Rossi F., Kerren A.}:
\newblock The state of the art in enhancing trust in machine learning models with the use of visualizations.
\newblock \emph{Computer Graphics Forum (CGF) 39}, 3 (2020), 713--756.

\bibitem[CMS99]{card1999}
\textsc{Card S.~K., Mackinlay J.~D., Shneiderman B.}:
\newblock \emph{Readings in information visualization - using vision to think}.
\newblock Academic Press, 1999.

\bibitem[CSS00]{chipman2000introduction}
\textsc{Chipman S.~F., Schraagen J.~M., Shalin V.~L.}:
\newblock Introduction to cognitive task analysis.
\newblock In \emph{Cognitive task analysis}. Psychology Press, 2000, pp.~17--38.

\bibitem[CT05]{cook2005illuminating}
\textsc{Cook K.~A., Thomas J.~J.}:
\newblock \emph{Illuminating the path: The research and development agenda for visual analytics}.
\newblock Tech. rep., Pacific Northwest National Lab.(PNNL), Richland, WA (United States), 2005.

\bibitem[DK23]{d2023data}
\textsc{D'ignazio C., Klein L.~F.}:
\newblock \emph{Data feminism}.
\newblock MIT press, 2023.

\bibitem[EBJ{\etalchar{*}}22]{tvcg2022eirich}
\textsc{Eirich J., Bonart J., J{\"{a}}ckle D., Sedlmair M., Schmid U., Fischbach K., Schreck T., Bernard J.}:
\newblock {IRVINE:} {A} design study on analyzing correlation patterns of electrical engines.
\newblock \emph{IEEE Transactions on Visualization and Computer Graphics (TVCG) 28}, 1 (2022), 11--21.

\bibitem[EFN12a]{endertFN12}
\textsc{Endert A., Fiaux P., North C.}:
\newblock Semantic interaction for sensemaking: Inferring analytical reasoning for model steering.
\newblock \emph{IEEE Trans. Visualization \& Computer Graphics (TVCG) 18}, 12 (2012), 2879--2888.

\bibitem[EFN12b]{endertFN12Text}
\textsc{Endert A., Fiaux P., North C.}:
\newblock Semantic interaction for visual text analytics.
\newblock In \emph{ACM Conf. Human Factors in Computing Systems (CHI)} (2012), {ACM}, pp.~473--482.

\bibitem[EHR{\etalchar{*}}14]{endertHRNFA14}
\textsc{Endert A., Hossain M.~S., Ramakrishnan N., North C., Fiaux P., Andrews C.}:
\newblock The human is the loop: new directions for visual analytics.
\newblock \emph{J. Intell. Inf. Syst. 43}, 3 (2014), 411--435.

\bibitem[ERT{\etalchar{*}}17]{endertRTWNBR17}
\textsc{Endert A., Ribarsky W., Turkay C., Wong B. L.~W., Nabney I.~T., Blanco I.~D., Rossi F.}:
\newblock The state of the art in integrating machine learning into visual analytics.
\newblock \emph{Computer Graphics Forum (CGF) 36}, 8 (2017), 458--486.

\bibitem[FDKN24]{fragiadakis2024}
\textsc{Fragiadakis G., Diou C., Kousiouris G., Nikolaidou M.}:
\newblock Evaluating human-ai collaboration: {A} review and methodological framework.
\newblock \emph{CoRR abs/2407.19098} (2024).

\bibitem[FOJ03]{fails2003interactive}
\textsc{Fails J.~A., Olsen~Jr D.~R.}:
\newblock Interactive machine learning.
\newblock In \emph{International Conference on Intelligent user interfaces} (2003), pp.~39--45.

\bibitem[FPSS96]{fayyad1996data}
\textsc{Fayyad U., Piatetsky-Shapiro G., Smyth P.}:
\newblock From data mining to knowledge discovery in databases.
\newblock \emph{AI magazine 17}, 3 (1996), 37--37.

\bibitem[FWR{\etalchar{*}}17]{federico2017}
\textsc{Federico P., Wagner M., Rind A., Amor{-}Amoros A., Miksch S., Aigner W.}:
\newblock The role of explicit knowledge: {A} conceptual model of knowledge-assisted visual analytics.
\newblock In \emph{IEEE Conf. Visual Analytics Science \& Technology (VAST)} (2017), {IEEE} Computer Society, pp.~92--103.

\bibitem[GT17]{gurel2017swot}
\textsc{Gürel E., Tat M.}:
\newblock Swot analysis: A theoretical review.
\newblock \emph{Journal of International Social Research 10}, 51 (2017), 994--1006.

\bibitem[HE24]{holterE24}
\textsc{Holter S., El{-}Assady M.}:
\newblock Deconstructing human-ai collaboration: Agency, interaction, and adaptation.
\newblock \emph{Computer Graphics Forum (CGF) 43}, 3 (2024).

\bibitem[Hoh21]{hohman21}
\textsc{Hohman F.}:
\newblock \emph{Interactive Scalable Interfaces for Machine Learning Interpretability}.
\newblock PhD thesis, Georgia Institute of Technology, Atlanta, GA, {USA}, 2021.

\bibitem[Hor99]{horvitz99}
\textsc{Horvitz E.}:
\newblock Principles of mixed-initiative user interfaces.
\newblock In \emph{ACM Conf. Human Factors in Computing Systems (CHI)} (1999), Williams M.~G., Altom M.~W., (Eds.), {ACM}, pp.~159--166.

\bibitem[JZF{\etalchar{*}}09]{jeongZFRC09}
\textsc{Jeong D.~H., Ziemkiewicz C., Fisher B.~D., Ribarsky W., Chang R.}:
\newblock ipca: An interactive system for pca-based visual analytics.
\newblock \emph{Computer Graphics Forum (CGF) 28}, 3 (2009), 767--774.

\bibitem[KAF{\etalchar{*}}08]{Keim2008}
\textsc{Keim D., Andrienko G., Fekete J.-D., G{\"o}rg C., Kohlhammer J., Melan{\c{c}}on G.}:
\newblock \emph{Visual Analytics: Definition, Process, and Challenges}.
\newblock Springer Berlin Heidelberg, Berlin, Heidelberg, 2008, pp.~154--175.

\bibitem[KCM89]{klein1989critical}
\textsc{Klein G.~A., Calderwood R., Macgregor D.}:
\newblock Critical decision method for eliciting knowledge.
\newblock \emph{IEEE Transactions on systems, man, and cybernetics 19}, 3 (1989), 462--472.

\bibitem[Kit14]{kitchin2014data}
\textsc{Kitchin R.}:
\newblock \emph{The data revolution: Big data, open data, data infrastructures and their consequences}.
\newblock Sage, 2014.

\bibitem[KLM{\etalchar{*}}08]{kehrerLMDSH08}
\textsc{Kehrer J., Ladst{\"{a}}dter F., Muigg P., Doleisch H., Steiner A.~K., Hauser H.}:
\newblock Hypothesis generation in climate research with interactive visual data exploration.
\newblock \emph{IEEE Trans. Visualization \& Computer Graphics (TVCG) 14}, 6 (2008), 1579--1586.

\bibitem[KTST15]{koh2015tools}
\textsc{KOH D. S.~M., TEE H.~G., SOH B.~K., TAN A. L.~S.}:
\newblock Tools for facilitating critical decision method during tacit knowledge elicitation.
\newblock In \emph{International Conference on Naturalistic Decision Making} (2015).

\bibitem[LAML22]{lin2022data}
\textsc{Lin H., Akbaba D., Meyer M., Lex A.}:
\newblock Data hunches: Incorporating personal knowledge into visualizations.
\newblock \emph{IEEE Transactions on Visualization and Computer Graphics 29}, 1 (2022), 504--514.

\bibitem[Lup17]{lupi2017data}
\textsc{Lupi G.}:
\newblock Data humanism: the revolutionary future of data visualization.
\newblock \emph{Print Magazine 30}, 3 (2017).

\bibitem[Mar95]{marchionini1995information}
\textsc{Marchionini G.}:
\newblock \emph{Information seeking in electronic environments}.
\newblock No.~9. Cambridge university press, 1995.

\bibitem[MD20]{meyerD20}
\textsc{Meyer M., Dykes J.}:
\newblock Criteria for rigor in visualization design study.
\newblock \emph{IEEE Trans. Visualization \& Computer Graphics (TVCG) 26}, 1 (2020), 87--97.

\bibitem[MH98]{militello1998applied}
\textsc{Militello L.~G., Hutton R.~J.}:
\newblock Applied cognitive task analysis (acta): a practitioner's toolkit for understanding cognitive task demands.
\newblock \emph{Ergonomics 41}, 11 (1998), 1618--1641.

\bibitem[MHH{\etalchar{*}}14]{mortierHHMC14}
\textsc{Mortier R., Haddadi H., Henderson T., McAuley D., Crowcroft J.}:
\newblock Human-data interaction: The human face of the data-driven society.
\newblock \emph{CoRR abs/1412.6159} (2014).

\bibitem[Mun09]{munzner09}
\textsc{Munzner T.}:
\newblock A nested process model for visualization design and validation.
\newblock \emph{IEEE Trans. Visualization \& Computer Graphics (TVCG) 15}, 6 (2009), 921--928.

\bibitem[NT95]{1995_Noanaka_Takeuchi}
\textsc{Nonaka I., Takeuchi H.}:
\newblock \emph{The knowledge-creating company: How japanese companies create the dynamics of innovation}.
\newblock Oxford University Press, New York, 1995.

\bibitem[OM20]{oppermannM20}
\textsc{Oppermann M., Munzner T.}:
\newblock Data-first visualization design studies.
\newblock In \emph{{IEEE} Workshop on Evaluation and Beyond - Methodological Approaches to Visualization ({BELIV})} (2020), Bezerianos A., Hall K.~W., Huron S., Kay M., Meyer M., Sedlmair M., (Eds.), {IEEE}, pp.~74--80.

\bibitem[OP10]{osterwalder2010business}
\textsc{Osterwalder A., Pigneur Y.}:
\newblock \emph{Business Model Generation: A Handbook for Visionaries, Game Changers, and Challengers}.
\newblock John Wiley \& Sons, 2010.

\bibitem[OPBS14]{osterwalder2014value}
\textsc{Osterwalder A., Pigneur Y., Bernarda G., Smith A.}:
\newblock \emph{Value Proposition Design: How to Create Products and Services Customers Want}.
\newblock John Wiley \& Sons, 2014.

\bibitem[PC05]{pirolli2005sensemaking}
\textsc{Pirolli P., Card S.}:
\newblock The sensemaking process and leverage points for analyst technology as identified through cognitive task analysis.
\newblock In \emph{Proceedings of international conference on intelligence analysis} (2005), vol.~5, McLean, VA, USA, pp.~2--4.

\bibitem[RAB{\etalchar{*}}24]{rogersABCFKKKTW24}
\textsc{Rogers J., Anastacio M., Bernard J., Chakhchoukh M., Faust R., Kerren A., Koch S., Kotthoff L., Turkay C., Wall E.}:
\newblock Visualization and automation in data science: Exploring the paradox of humans-in-the-loop.
\newblock In \emph{{IEEE} Visualization in Data Science (VDS)} (2024), {IEEE}, pp.~1--5.

\bibitem[RBB{\etalchar{*}}23]{rosaBBASCBGA23}
\textsc{Rosa B.~L., Blasilli G., Bourqui R., Auber D., Santucci G., Capobianco R., Bertini E., Giot R., Angelini M.}:
\newblock State of the art of visual analytics for explainable deep learning.
\newblock \emph{Computer Graphics Forum (CGF) 42}, 1 (2023), 319--355.

\bibitem[Rie19]{riedl2019}
\textsc{Riedl M.~O.}:
\newblock Human-centered artificial intelligence and machine learning.
\newblock \emph{Human Behavior and Emerging Technologies 1}, 1 (2019), 33--36.

\bibitem[RM23]{rezwanaM23}
\textsc{Rezwana J., Maher M.~L.}:
\newblock Designing creative {AI} partners with {COFI:} {A} framework for modeling interaction in human-ai co-creative systems.
\newblock \emph{{ACM} Trans. Comput. Hum. Interact. 30}, 5 (2023), 67:1--67:28.

\bibitem[RMS{\etalchar{*}}20]{ramos2020interactive}
\textsc{Ramos G., Meek C., Simard P., Suh J., Ghorashi S.}:
\newblock Interactive machine teaching: a human-centered approach to building machine-learned models.
\newblock \emph{Human--Computer Interaction 35}, 5-6 (2020), 413--451.

\bibitem[Rus22]{russell2022}
\textsc{Russell S.}:
\newblock Artificial intelligence and the problem of control.
\newblock In \emph{Perspectives on Digital Humanism}. 2022, pp.~19--24.

\bibitem[Shn96]{shneiderman96}
\textsc{Shneiderman B.}:
\newblock The eyes have it: {A} task by data type taxonomy for information visualizations.
\newblock In \emph{{IEEE} Symposium on Visual Languages} (1996), {IEEE} Computer Society, pp.~336--343.

\bibitem[Shn21]{shneiderman21}
\textsc{Shneiderman B.}:
\newblock Tutorial: Human-centered {AI:} reliable, safe and trustworthy.
\newblock In \emph{International Conference on Intelligent User Interfaces} (2021), Hammond T., Verbert K., Parra D., (Eds.), {ACM}, pp.~7--8.

\bibitem[Shn22]{shneiderman22}
\textsc{Shneiderman B.}:
\newblock Human-centered {AI:} ensuring human control while increasing automation.
\newblock In \emph{Workshop on Human Factors in Hypertext (HUMAN} (2022), {ACM}, pp.~1:1--1:2.

\bibitem[SJB{\etalchar{*}}21]{cag2021sperrle}
\textsc{Sperrle F., Jeitler A., Bernard J., Keim D., El-Assady M.}:
\newblock Co-adaptive visual data analysis and guidance processes.
\newblock \emph{Computers \& Graphics} (2021).

\bibitem[SKKC19]{sachaKKC19}
\textsc{Sacha D., Kraus M., Keim D.~A., Chen M.}:
\newblock {VIS4ML:} an ontology for visual analytics assisted machine learning.
\newblock \emph{IEEE Trans. Visualization \& Computer Graphics (TVCG) 25}, 1 (2019), 385--395.

\bibitem[SMM12]{sedlmairMM12}
\textsc{Sedlmair M., Meyer M.~D., Munzner T.}:
\newblock Design study methodology: Reflections from the trenches and the stacks.
\newblock \emph{IEEE Trans. Visualization \& Computer Graphics (TVCG) 18}, 12 (2012), 2431--2440.

\bibitem[SSS{\etalchar{*}}14]{sachaKGM2014}
\textsc{{Sacha} D., {Stoffel} A., {Stoffel} F., {Kwon} B.~C., {Ellis} G., {Keim} D.~A.}:
\newblock Knowledge generation model for visual analytics.
\newblock \emph{IEEE Trans. Visualization \& Computer Graphics (TVCG) 20}, 12 (2014), 1604--1613.

\bibitem[SSSE20]{spinnerSSE20}
\textsc{Spinner T., Schlegel U., Sch{\"{a}}fer H., El{-}Assady M.}:
\newblock explainer: {A} visual analytics framework for interactive and explainable machine learning.
\newblock \emph{IEEE Trans. Visualization \& Computer Graphics (TVCG) 26}, 1 (2020), 1064--1074.

\bibitem[SSZ{\etalchar{*}}16]{sachaSZLWNK16}
\textsc{Sacha D., Sedlmair M., Zhang L., Lee J.~A., Weiskopf D., North S.~C., Keim D.~A.}:
\newblock Human-centered machine learning through interactive visualization: review and open challenges.
\newblock In \emph{European Symposium on Artificial Neural Networks6} (2016).

\bibitem[SSZ{\etalchar{*}}17]{sachaSZLPWNK17}
\textsc{Sacha D., Sedlmair M., Zhang L., Lee J.~A., Peltonen J., Weiskopf D., North S.~C., Keim D.~A.}:
\newblock What you see is what you can change: Human-centered machine learning by interactive visualization.
\newblock \emph{Neurocomputing 268} (2017), 164--175.

\bibitem[WBV22]{wondimu2022}
\textsc{Wondimu N.~A., Buche C., Visser U.}:
\newblock Interactive machine learning: {A} state of the art review.
\newblock \emph{CoRR abs/2207.06196} (2022).

\bibitem[WH00]{wirth2000crisp}
\textsc{Wirth R., Hipp J.}:
\newblock Crisp-dm: Towards a standard process model for data mining.
\newblock In \emph{Proceedings of the 4th international conference on the practical applications of knowledge discovery and data mining} (2000), vol.~1, Manchester, pp.~29--39.

\bibitem[WJD{\etalchar{*}}09]{wangJDLRC09}
\textsc{Wang X., Jeong D.~H., Dou W., Lee S., Ribarsky W., Chang R.}:
\newblock Defining and applying knowledge conversion processes to a visual analytics system.
\newblock \emph{Comput. Graph. 33}, 5 (2009), 616--623.

\end{thebibliography}

\end{document}